\documentclass[pre,twocolumn,amsmath,amssymb,showpacs]{revtex4}

\usepackage{graphicx}
\usepackage{amssymb}
\usepackage{epstopdf}
\usepackage{times}
\usepackage{dcolumn}
\usepackage{bm}
\usepackage{amsmath}
\usepackage[dvips]{color}

\newcommand{\bfa}{\mbox{\boldmath $\it a$}}
\newcommand{\bfb}{\mbox{\boldmath $\it b$}}  
 
\newcommand{\bfX}{\mbox{\boldmath $\it X$}} 
\newcommand{\caJ}{\mathcal{J}} 
\newcommand{\caL}{\mathcal{L}} 
\newcommand{\caO}{\mathcal{O}} 
\newcommand{\caP}{\mathcal{P}} 

\newcommand{\Real}[1]{ \mbox{Re}\!\left[ #1 \right] } 
\newcommand{\Imaginary}[1]{ \,\mbox{Im}\!\left[ #1 \right] }  
\newcommand{\InPro}[2]{\left\langle \,#1 \;\right|\left. #2 \,\right\rangle}   
\newcommand{\spaEq}{\hspace{0.3cm}}   
\newcommand{\spaNeg}{\hspace{-0.6cm}}   

\newcommand{\vspfigA}{\vspace{0.2cm}}  
\newcommand{\vspfigB}{\vspace{0.1cm}} 
\newcommand{\vspfigC}{\vspace{0.3cm}}
\newcommand{\widthfig}{0.46\textwidth}

\begin{document}

\baselineskip 0.415cm
\title{Particle escapes in an open quantum network via multiple leads}

\author{Tooru Taniguchi  and Shin-ichi Sawada}

\affiliation{School of Science and Technology, Kwansei Gakuin University,
2-1 Gakuen, Sanda city, Japan} 

\date{\today}
\begin{abstract}

Quantum escapes of a particle from an end of a one-dimensional finite region to $N$ number of semi-infinite leads are discussed by a scattering theoretical approach. Depending on a potential barrier amplitude at the junction, the probability $P(t)$ for a particle to remain in the finite region at time $t$ shows two deferent decay behaviors after a long time; one is proportional to $N^{2}/t^{3}$ and another is proportional to $1/(N^{2}t)$. In addition, the velocity $V(t)$ for a particle to leave from the finite region, defined from a probability current of the particle position, decays in power  $\sim 1/t$ asymptotically in time, independently of the number $N$ of leads and the initial wave function, etc. For a finite time, the probability $P(t)$ decays exponentially in time with a smaller decay rate for more number $N$ of leads, and the velocity $V(t)$ shows a time-oscillation whose amplitude is larger for more number $N$ of leads. Particle escapes from the both ends of a finite region to multiple leads are also discussed by using a different boundary condition. 

\end{abstract}


\maketitle

\section{Introduction} 

   The particle escape is a typical nonequilibrium phenomenon in open systems. 
   It is a current of particles from a region where the particles were initially confined. 
   The concept of escape has been used to describe a variety of physical phenomena, such as $\alpha$-decaying nucleus \cite{GC29,W61,DN02}, chemical reactions as Kramers' escape problem \cite{K40,HTB90,K92}, etc.  
   Some dynamical properties, like chaos \cite{BB90,AGH96,PB00}, the recurrence time \cite{AT09}, the first-passage time \cite{HTB90,K92}, transport coefficients \cite{GN90,G98,K07}, etc., have also been investigated via escape behaviors of particles. 
   The particle escapes were discussed in many types of systems, for example, stochastic systems \cite{K40,HTB90,K92}, classical billiard systems \cite{BB90,AGH96,MHC01,FKC01,BD07}, map systems \cite{PB00,AT09,DY06} and quantum systems \cite{GC29,W61,DN02,LW91,CM97,ZB03}. 

   Escape phenomena in open systems cause a decay of various quantities by a particle current from a finite region. 
   For instance, the probability for particles to remain in the initially confined region, which we call the survival probability in this paper, would decay in time, if particles continue to be flowed out from the initial region. 
   Such decay properties in escape systems have led to some interesting results and conjectures. 
   For example, in classical mechanical systems with a particle escape via a small hole, the survival probability would decay exponentially in time if the dynamics is chaotic, while it would decay in power for non-chaotic systems \cite{BB90,AGH96,PB00,AT09,FKC01}. 
   On the other hand, in many quantum mechanical systems, the survival probability decays in a power asymptotically in time \cite{DN02,LW91,ZB03,GMV07,OK91,DHM92} with an exponential decay for a finite time \cite{DN02,ZB03,GMV07,OK91}, and values of the power in the decay vary by initial conditions \cite{M03} or particle interactions \cite{TS11,C11}, etc. 

   In this paper we discuss particle escapes in quantum mechanical networks as an example of open dynamical systems. 
   The quantum network system is also called the ``quantum graph,'' and is constructed by connecting finite and infinite narrow wires like a network, and have been widely used as models to describe mesoscopic transports like Aharonov-Bohm type of effects \cite{BI84,B85}, resonance tunnellings \cite{PS92,TB99}, current splitters \cite{ES88,BJ05,SM10}, chaos and diffusion \cite{KS00,BG01}, etc. 
   Steady electric currents in open quantum network systems are described by quantum scattering theory \cite{AS91,KS99,T01,TM01}.
   This kind of quantum systems with narrow wires could be experimentally realized as a combination of atomic or molecular wires or as a graph-like structure on the surface of a semiconductor by a recent development of nano-technology \cite{D95,I97,B09}. 

   An important feature of network systems is an effect of current splitter at a network junction. 
   In order to consider such a splitting effect of currents in quantum escapes as simple and concrete as possible, we consider particle escapes from a finite one-dimensional region with the length $L$ via $N$-number of semi-infinite one-dimensional leads.  
   The multiple leads is connected at one end of the finite region with a potential barrier amplitude $\Lambda$ at the junction, and we impose the fixed boundary condition at another terminated end of the finite region. 
   As a theoretical approach to describe particle escapes in such a quantum network, we use a quantum scattering theoretical approach, by which we consider concretely decay properties of two quantities to characterize the quantum escapes.
   The first quantity of decaying in the particle escape is the survival probability $P(t)$ for the particle to remain in the finite region at time $t$. 
   We show analytically that the survival probability $P(t)$ depends on the number $N$ of attached semi-infinite leads as the $N^{2}$ proportional ratio, i.e. $\lim_{t\rightarrow +\infty}P(t)/[P(t)|_{N=1}]=N^{2}$, in the case of $\Lambda  \neq -\hbar^{2}/(2mL)$ ($m$: the mass of particle, $2\pi \hbar$: the Planck constant). 
   This means that after a long time the particle can remain more inside the original finite region by connecting more semi-infinite leads. 
   Besides, in this case the survival probability $P(t)$ decays in power $\sim 1/t^{3}$ asymptotically in time. 
   In contrast, in the case of $\Lambda = -\hbar^{2}/(2mL)$ we obtain the $N^{2}$ inversely proportional ratio $\lim_{t\rightarrow +\infty}P(t)/[P(t)|_{N=1}]=1/N^{2}$ for the survival probability $P(t)$, meaning that after a long time a particle escapes more by connecting more leads. 
   Here, the survival probability $P(t)$ decays in power $\sim 1/t$ asymptotically in time, differently from the case of $\Lambda  \neq -\hbar^{2}/(2mL)$.   
   We also discuss finite time properties of the survival probability numerically, and show that for a finite time the survival probability decays exponentially for a longer time with a smaller decay rate by connecting more number $N$ of leads.  
   As the second quantity to investigate decay properties in quantum particle escapes, we consider  the velocity $V(t)$ for the particle to escape from the finite region, which is introduced by the equation of continuity for the particle position probability. 
   We show analytically that the escape velocity $V(t)$ behaves asymptotically in time as $V(t) \;\overset{t\rightarrow+\infty}{\sim}\; L/t$ for $\Lambda  \neq -\hbar^{2}/(2mL)$ and $V(t) \;\overset{t\rightarrow+\infty}{\sim}\; L/(3t)$ for $\Lambda  = -\hbar^{2}/(2mL)$, which are independent of the number $N$ of leads and the initial wave function, etc. 
   It is also shown by numerical calculations that for a finite time the escape velocity $V(t)$ oscillates in time, and takes a larger magnitude of the oscillations for more number $N$ of attached leads. 
   Furthermore, by using a different boundary condition, our scattering approach to quantum network systems also allow us to discuss particle escapes from a finite region with the length $2L$ whose both end are connected to $N$ number of semi-infinite one-dimensional leads. 
   
   The outline of this paper is as follows. 
   In Sect. \ref{QuantumScatteringTheory}, the quantum network system with a finite wire connected to multiple leads is introduced. 
   Based on the equation of continuity for particle position probability, we introduce a probability current whose conservation imposes boundary conditions at the junction of leads and at the terminated end of the finite wire. 
   These boundary conditions specify the scattering states of this system, from which we describe the time-evolution of a wave function of the system. 
   In Sect. \ref{EscapeBehavior}, we introduce the survival probability $P(t)$ from the particle position probability and the escape velocity $V(t)$ from the probability current in the quantum network system, and discuss these decay properties.  
   Finally, we give some conclusion and remarks in Sect. \ref{ConclusionRemarks}.

\section{Quantum scattering approach to network systems with multiple leads} 
\label{QuantumScatteringTheory}

\subsection{Quantum network system and boundary conditions}
\label{NetworkSystemBoundary}

\begin{figure}[!t]
\vspfigA
\begin{center}
\includegraphics[width=\widthfig]{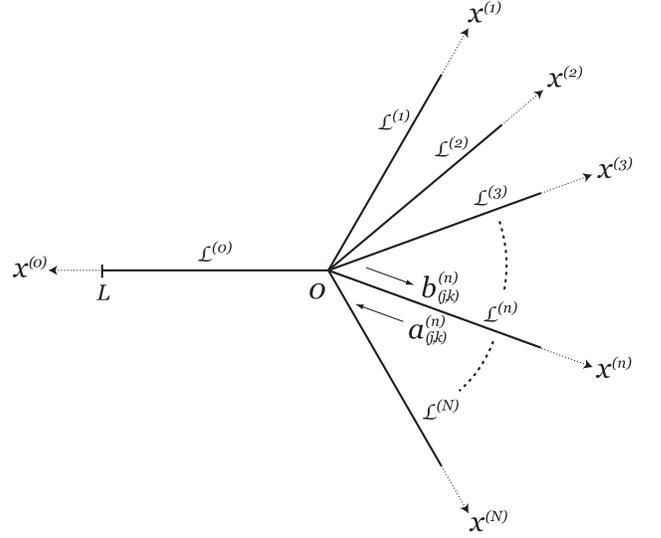}
\vspfigB
\caption{
   A quantum network with a junction for multiple leads. 
   It consists of a finite region $\caL^{(0)}$ with a length $L$ and the semi-infinite one-dimensional region $\caL^{(n)}$ ($n=1,2,\cdots,N$) connected to an end of the region $\caL^{(0)}$. 
   The $x^{(n)}$-axis of coordinates is taken in the region $\caL^{(n)}$ with the origin $O$ at the junction and its coordinate satisfies $x^{(n)}>0$ ($n=0,1,2,\cdots,N$). 
   The coefficients $a_{(j,k)}^{(n)}$ ($b_{(j,k)}^{(n)}$) is the amplitude of the incident wave to  (the scattered wave from) the junction in the region $\caL^{(n)}$ by a plain wave injected from the $j$-th lead with the wave number $k$. 
}
\label{Fig1SystemA}
\end{center}
\vspfigC
\end{figure}

   We consider quantum network systems consisting of a finite one-dimensional segment with a length $L$ whose end is connected to $N$ number of semi-infinite one-dimensional leads. 
   (See Fig. \ref{Fig1SystemA} as a schematic illustration of this network.)  
   We call the finite segment with the the length $L$ the region $\caL^{(0)}$ and also call the $j$-th semi-infinite segment of lead the region $\caL^{(n)}$ ($n=1,2,\cdots,N$). 
   In each region $\caL^{(n)}$ we put the $x^{(n)}$-axis of coordinates with the origin $O$ at the junction of leads, in which the positive direction of the $x^{(n)}$-axis is taken from the origin $O$ to the region $\caL^{(n)}$ ($n=0,1,2,\cdots,N$).  
%

   We introduce the wave function $\Psi^{(n)}(x,t)$ of a particle in this quantum network at  time $t$ and the position $x\in \caL^{(n)}$ for $x>0$. 
   From Schr\"odinger equation for the wave function $\Psi^{(n)}(x,t)$ with a real potential we derive the equation of continuity for the particle position probability density
\begin{eqnarray}
   \rho^{(n)}(x,t) \equiv |\Psi^{(n)}(x,t)|^{2}
\label{ProbaDensi1}
\end{eqnarray}
as 
\begin{eqnarray}
   \frac{\partial \rho^{(n)}(x,t) }{\partial t} + \frac{\partial \rho^{(n)}(x,t) v^{(n)}(x,t)}{\partial x} = 0
\end{eqnarray}
in which the local velocity $v^{(n)}(x,t)$ at the position $x\in \caL^{(n)}$ and the time $t$ is introduced as  
\begin{eqnarray}
   v^{(n)}(x,t) 
   \equiv \frac{\hbar}{m}
      \Imaginary{\frac{\frac{\partial \Psi^{(n)}(x,t)}{\partial x}}{\Psi^{(n)}(x,t)}}
\label{LocalVeloc1}
\end{eqnarray}
with the mass $m$ of particle and the Planck constant $2\pi\hbar$ \cite{Memo3}. 
   Here, $\Imaginary{X}$ means the imaginary part of $X$ for any complex number $X$.  
   
For describing the quantum state at the junction $O$, we impose that the wave function of the system is continuous at any position, including at the origin $O$ so that the boundary conditions
\begin{eqnarray}
   \Psi(0,t) 
   &\equiv&  
   \lim_{x\rightarrow +0}\Psi^{(0)}(x,t) 
   = \lim_{x\rightarrow +0}\Psi^{(1)}(x,t)
      \nonumber \\
   &=& \cdots 
   = \lim_{x\rightarrow +0}\Psi^{(N)}(x,t)
\label{BoundPosit1}
\end{eqnarray}
are satisfied at any time $t$. 
   We further assume that there is no net particle current source at the junction $O$, namely  $ \lim_{x\rightarrow +0} \sum_{n=0}^{N} $ $\rho^{(n)}(x,t) v^{(n)}(x,t) = 0$, leading to 
\begin{eqnarray}
   \lim_{x\rightarrow +0} \sum_{n=0}^{N} v^{(n)}(x,t) = 0 
\label{BoundMomen1}
\end{eqnarray}
for the local velocity (\ref{LocalVeloc1}), noting that by the condition (\ref{BoundPosit1}) the position probability density $\lim_{x\rightarrow +0}\rho^{(n)}(x,t)$ is independent of the region number $n$ \cite{Memo5}. 
   Using Eqs.  (\ref{LocalVeloc1}) and (\ref{BoundPosit1}), Eq. (\ref{BoundMomen1}) is rewritten as 
\begin{eqnarray}
   \lim_{x\rightarrow +0} \sum_{n=0}^{N}\frac{\partial \Psi^{(n)}(x,t)}{\partial x} 
   = \lambda \Psi(0,t)
\label{BoundVeloc1}
\end{eqnarray}
with a real constant $\lambda$ \cite{AS91,TM01,Memo7}. 
   It is important to note that for the case of $N=1$ the condition (\ref{BoundVeloc1}) is equivalent to the boundary condition for a delta functional potential with the amplitude $\hbar^{2}\lambda/(2m)$ in an one-dimensional one-particle system. 
   In this sense, we regard the real constant
\begin{eqnarray}
   \Lambda\equiv\frac{\hbar^{2}\lambda}{2m}
\label{AmpliPoten1}
\end{eqnarray}
with the constant $\lambda$ appearing in Eq. (\ref{BoundVeloc1}) as a potential barrier amplitude at the junction $O$. 
   Finally, we impose that there is  no particle current source at the end $x^{(0)}=L$ of the region $\caL^{(0)}$, so
\begin{eqnarray}
   v^{(0)}(L,t) = 0, 
\label{BoundMomen2a}
\end{eqnarray}
namely 
\begin{eqnarray}
    \left.\frac{\partial \Psi^{(0)}(x,t)}{\partial x} \right|_{x=L}
   = \mu \Psi^{(0)}(L,t)
\label{BoundMomen2}
\end{eqnarray}
with a real constant $\mu$, similarly to Eq. (\ref{BoundVeloc1}) \cite{Memo5,Memo7}.

\subsection{Scattering states and matrix} 

   We assume that the quantum network introduced in the previous subsection \ref{NetworkSystemBoundary} consists of ideal leads, i.e. there is no potential in any of the region $\caL^{(n)}$, $n=0,1,\cdots,N$ except at the junction. 
   In this case, the energy eigenstate $\Phi_{(j,k)}^{(n)}(x)$ at the point $x$ in $\caL^{(n)}$ is presented by a superposition of plain waves with the wave number $k \;(>0)$ as 
\begin{eqnarray}
   \Phi_{(j,k)}^{(n)}(x) = a_{(j,k)}^{(n)} e^{-ikx} + b_{(j,k)}^{(n)} e^{ikx} ,
\label{ScattState1}
\end{eqnarray}
$n=0,1,\cdots,N$, corresponding to the energy eigenvalue $E_{k}=\hbar^{2}k^{2}/(2m)$. 
   Here, we introduced the suffix $j$ in the quantities $\Phi_{(j,k)}^{(n)}(x)$, $a_{(j,k)}^{(n)}$ and $b_{(j,k)}^{(n)}$ to distinguish different states with the same energy $E_{k}$ as discussed later in detail. 
   The constants $a_{(j,k)}^{(n)}$ and $b_{(j,k)}^{(n)}$ in Eq. (\ref{ScattState1}) are regarded as the amplitude of incident wave to and scattered wave from the junction $O$, respectively. 
   The $(N+1)$ dimensional column vector $\bfb_{(j,k)}\equiv(b_{(j,k)}^{(0)} b_{(j,k)}^{(1)} \cdots b_{(j,k)}^{(N+1)})^{T}$ is connected to the $(N+1)$ dimensional column vector $\bfa_{(j,k)}\equiv(a_{(j,k)}^{(0)} a_{(j,k)}^{(1)} \cdots a_{(j,k)}^{(N+1)})^{T}$ as 
\begin{eqnarray}
   \bfb_{(j,k)} = S_{k} \bfa_{(j,k)}
\label{ScattMatri1}
\end{eqnarray}
with the scattering matrix $S_{k}$. 
   Here, the notation $\bfX^{T}$ denotes the transpose of $\bfX$ for any vector $\bfX$.
   In Eq. (\ref{ScattMatri1}) we suppressed the suffix $j$ for the scattering matrix $S_{k}$, because as shown later in Eq. (\ref{ScattMatri2}) the scattering matrix is independent of the suffix $j$.
   The energy eigenstate (\ref{ScattState1}), which has nonzero value even in the infinite region $x^{(n)}\rightarrow +\infty$ of $\caL^{(n)}$, $n=1,2,\cdots,N$, is the scattering state of the quantum network system. 

   The scattering state (\ref{ScattState1})  as a special example of the wave function $\Psi^{(n)}(x,t)$ must satisfy the conditions (\ref{BoundPosit1}) and (\ref{BoundVeloc1}), leading to the specific form of the scattering matrix $S_{k}$ as
\begin{eqnarray}
   S_{k}=\frac{2}{N+1+i\frac{\lambda}{k}} U-I.
\label{ScattMatri2}
\end{eqnarray}
Here, $I$ is  the $(N+1)\times(N+1)$ identical matrix, and $U$ is the $(N+1)\times(N+1)$ matrix whose all elements are given by $1$. 
   [See Appendix \ref{PropeScattMatri} for an derivation of Eq. (\ref{ScattMatri1}) with the scattering matrix (\ref{ScattMatri2}). ] 
   Noting the relation $U^{2}=(N+1)U$, the scattering matrix (\ref{ScattMatri2}) is shown to be an unitary matrix satisfying the relation 
\begin{eqnarray}
   S_{k}^{\dagger}S_{k} = S_{k}S_{k}^{\dagger} = I
\label{UnitaScatt1}
\end{eqnarray}
with the Hermitian matrix $S_{k}^{\dagger}$ of $S_{k}$, supporting the relation $|\bfb_{(j,k)}|^{2} = |\bfa_{(j,k)}|^{2}$ by Eq. (\ref{ScattMatri1}). 
      
\begin{figure}[!t]
\vspfigA
\begin{center}
\includegraphics[width=\widthfig]{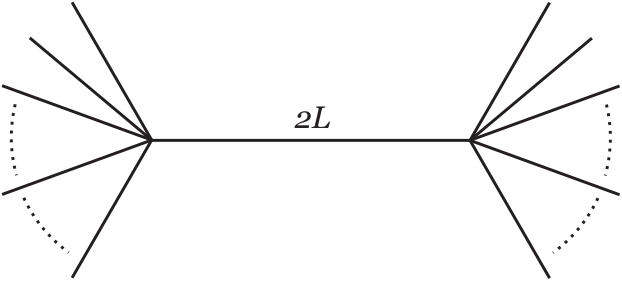}
\vspfigB
\caption{
   A quantum network with two junctions for multiple leads at the both ends of the finite region with a length $2L$. 
}
\label{Fig1SystemB}
\end{center}
\vspfigC
\end{figure}

   On the other hand, the condition (\ref{BoundMomen2}) leads to the relation 
\begin{eqnarray}
   a_{(j,k)}^{(0)} = \frac{1+i\frac{\mu}{k}}{1-i\frac{\mu}{k}}e^{2ikL} b_{(j,k)}^{(0)}
\label{ScattMatri3}
\end{eqnarray}
between the amplitudes $a_{(j,k)}^{(0)}$ and $b_{(j,k)}^{(0)}$ for the scattering state in the finite region $\caL^{(0)}$.  
   The condition (\ref{ScattMatri3}) guarantees the same magnitude $|b_{(j,k)}^{(0)}|^{2}=|a_{(j,k)}^{(0)}|^{2}$ of the plain waves propagating to the opposite directions with each other in the finite region $\caL^{(0)}$. 
   The relation (\ref{ScattMatri3}) includes two special cases: 
\begin{eqnarray}
   a_{(j,k)}^{(0)} &=& - e^{2ikL} b_{(j,k)}^{(0)} \;\;\;\;\;\; \mbox{for $\mu\rightarrow+\infty$},
      \\
   a_{(j,k)}^{(0)} &=& + e^{2ikL} b_{(j,k)}^{(0)}\;\;\;\;\;\; \mbox{for $\mu = 0$}, 
\end{eqnarray}
meaning that the limit $\mu\rightarrow+\infty$ imposes the fixed (Dirichlet) boundary condition $\Phi_{(j,k)}^{(0)}(L) = 0$ for scattering states and the case of $\mu=0$ imposes the open (Neuman) boundary condition at the end $x^{(0)}=L$ of the finite region $\caL^{(0)}$. 
   It is important to note that the open boundary case $\mu=0$ and the fixed boundary case $\mu\rightarrow+\infty$ can also describe states in the quantum network consisting of a finite lead with a length $2L$ whose {\em both} sides are connected to $N$ number of multiple semi-infinite leads, as shown in Fig. \ref{Fig1SystemB}.  
   Here, the open boundary case $\mu=0$ (the fixed boundary case $\mu\rightarrow+\infty$) corresponds to the quantum state described by the wave function with the inversion symmetry (anti-symmetry) at the center of the finite region with the length $2L$ in such a system. 
%

   By Eqs. (\ref{ScattMatri1}), (\ref{ScattMatri2}) and (\ref{ScattMatri3}) the coefficient $b_{(j,k)}^{(n)}$ is given by 
\begin{eqnarray}
   b_{(j,k)}^{(0)} &=& \frac{2\sum_{n'=1}^{N}a_{(j.k)}^{(n')}}
      {N+1+i\frac{\lambda}{k}+\left(N-1+i\frac{\lambda}{k}\right)
      \frac{1+i\frac{\mu}{k}}{1-i\frac{\mu}{k}}e^{2ikL}}, 
\label{ScattCoeff1}\\
   b_{(j,k)}^{(\eta)} &=& - a_{(j.k)}^{(\eta)} 
      \nonumber \\
      &&
      + \frac{2\left(1+\frac{1+i\frac{\mu}{k}}{1-i\frac{\mu}{k}}e^{2ikL}\right)
      \sum_{n'=1}^{N}a_{(j.k)}^{(n')}}
      {N+1+i\frac{\lambda}{k}+\left(N-1+i\frac{\lambda}{k}\right)
      \frac{1+i\frac{\mu}{k}}{1-i\frac{\mu}{k}}e^{2ikL}}
      \;\;\;\;\;\;
\label{ScattCoeff2}
\end{eqnarray}
for $\eta=1,\cdots,N$. 
   Eqs. (\ref{ScattMatri3}), (\ref{ScattCoeff1}) and  (\ref{ScattCoeff2}) shows that the coefficients $a_{(j,k)}^{(0)}$ and $b_{(j,k)}^{(n)}$, $n=0,\cdots,N$ are given from the coefficients $a_{(j,k)}^{(n)}$, $n=1,\cdots,N$. 
   We can still take any value for $a_{(j,k)}^{(n)}$, $n=1,\cdots,N$, as the $N$ number of incident wave amplitudes from infinite regions $x^{(n)}\rightarrow+\infty$, $n=1,\cdots,N$, and this voluntariness leads to necessity of the suffix $j$ to distinguish degeneracy of the energy eigenstates $\Phi_{(j,k)}^{(n)}(x)$ corresponding to the energy $E_{k}$.  
   In this paper, we introduce $a_{(j,k)}^{(n)}$ as 
\begin{eqnarray}
   a_{(j,k)}^{(n)} \equiv \varsigma_{j}\delta_{jn}
\label{ScattCoeff3}
\end{eqnarray}
for $n=1,2,\cdots,N$ with a constant $\varsigma_{j}$. 
   Eq. (\ref{ScattCoeff3}) means that the state $\Phi_{(j,k)}^{(n)}(x)$ is the scattering state produced by the incident wave from the infinite region $x^{(j)}\rightarrow +\infty$ of $\caL^{(j)}$ only. 
        
   We define the inner products
\begin{eqnarray}
   \InPro{X}{Y} &\equiv&  
      \lim_{\epsilon\rightarrow +0}\int_{\epsilon}^{L} dx \; X^{(0)}(x)^{*}Y^{(0)}(x)
      \nonumber \\
      &&
      + \lim_{\epsilon\rightarrow +0}\sum_{n=1}^{N}
      \int_{\epsilon}^{+\infty} dx \; X^{(n)}(x)^{*}Y^{(n)}(x) 
      \nonumber \\
   &=& \sum_{n=0}^{N} \int_{\caL^{(n)}} dx\;  X^{(n)}(x)^{*}Y^{(n)}(x) 
\label{InneProdu1}
\end{eqnarray}
for any quantum states $X$ and $Y$, whose values at the position $x$ in the region $\caL^{(n)}$ are given by $X^{(n)}(x)$ and $Y^{(n)}(x)$, respectively. 
   Here, $X^{*}$ with the asterisk denotes the complex conjugate of $X$ for any complex number $X$. 
   For this inner product  (\ref{InneProdu1}) it is essential to note that the scattering state $\Phi_{(j,k)}^{(n)}(x)$ satisfies the orthogonal relation 
\begin{eqnarray}
   \InPro{\Phi_{(j,k)}}{\Phi_{(j',k')}} = \delta_{jj'}\delta(k-k')
\label{OrthoScatt1}
\end{eqnarray}
for $kk'>0$. Here, we specified the constant $\varsigma_{j}$ in Eq. (\ref{ScattCoeff3}) as 
\begin{eqnarray}
   \varsigma_{j} \equiv \frac{1}{\sqrt{2\pi}} ,
\label{ConstA1}
\end{eqnarray}
independently with respect to the suffix $j = 1,2,\cdots,N$, so that the coefficient of the term $\delta_{jj'}\delta(k-k')$ in the right-hand side of Eq. (\ref{OrthoScatt1}) becomes $1$. 
   The proof of Eq. (\ref{OrthoScatt1}) is given in Appendix \ref{OrthoScattState}.

\subsection{Green function and the time-evolution of wave function} 

   Now, we assume that the initial quantum state $\Psi(0)$ at $t=0$ is expanded by the scattering states, as 
\begin{eqnarray}
   \Psi^{(n)}(x,0) = \sum_{j=1}^{N}\int_{0}^{+\infty} dk\; A_{(j,k)} \Phi_{(j,k)}^{(n)}(x)
\label{InitiState1}
\end{eqnarray}
at the position $x$ in $\caL^{(n)}$ with a constant $A_{(j,k)}$. 
   By Eqs. (\ref{InneProdu1}), (\ref{OrthoScatt1}) and (\ref{InitiState1}) the constant $A_{(j,k)}$ is expressed as 
\begin{eqnarray}
   A_{(j,k)} = \InPro{\Phi_{(j,k)}}{\Psi(0)} 
\label{CoeffA1}
\end{eqnarray}
for $k>0$. 
   The initial condition (\ref{InitiState1}) implies that the initial state $\Psi(0)$ does not include bound states with discretized energy spectrum which are orthogonal to the scattering states \cite{Memo1}.  

   The wave function of the system at time $t>0$ is given by applying the time-evolutional operator $\exp(-i\hat{H}t/\hbar)$ with the Hamiltonian operator $\hat{H}$ to the initial state (\ref{InitiState1}), and we obtain 
\begin{eqnarray}
   \Psi^{(n)}(x,t) &=& \sum_{j=1}^{N}\int_{0}^{+\infty} dk\; 
      A_{(j,k)} e^{-iE_{k}t/\hbar} \Phi_{(j,k)}^{(n)}(x)
      \nonumber \\
   &=& \sum_{n'=0}^{N}\int_{\caL^{(n')}} dx'\; G^{(n,n')}(x,x';t) 
   \Psi^{(n')}(x',0)
   \nonumber\\
\label{WaveFunct1}
\end{eqnarray}
by Eqs. (\ref{InneProdu1}), (\ref{CoeffA1}) and $\hat{H} \Phi_{(j,k)}^{(n)}(x) = E_{k} \Phi_{(j,k)}^{(n)}(x)$. 
   Here, $G^{(n,n')}(x,x';t)$ is the Green function defined as
\begin{eqnarray}
  && 
  G^{(n,n')}(x,x';t) 
     \nonumber \\
   &&\spaEq \equiv 
   \sum_{j=1}^{N}\int_{0}^{+\infty} dk\; 
      \Phi_{(j,k)}^{(n)}(x) 
   \Phi_{(j,k)}^{(n')}(x')^{*} e^{-iE_{k}t/\hbar} .
   \spaEq\spaEq
\label{GreenFunct1}
\end{eqnarray}
In other words, we can calculate the Green function (\ref{GreenFunct1}) via the scattering state (\ref{ScattState1}) with the coefficients (\ref{ScattMatri3}), (\ref{ScattCoeff1}), (\ref{ScattCoeff2}) and (\ref{ScattCoeff3}), leading to the wave function at time $t$ by Eq. (\ref{WaveFunct1}).

\section{Escape behavior of a particle via multiple leads} 
\label{EscapeBehavior}
\subsection{Quantum escapes in a network system} 

   In this section we consider behaviors of a particle to escape from the finite region $\caL^{(0)}$ to the semi-infinite regions $\caL^{(n)}$, $n=1,2,\cdots,N$. 
   In order to discuss such phenomena, we set the initial wave function at $t=0$ to take nonzero value only in the finite region $\caL^{(0)}$, i.e. 
\begin{eqnarray}
   \Psi^{(n)}(x,0) = 0
      \;\;\; \mbox{for $n=1,\cdots,N$} .
\label{InitiState2}
\end{eqnarray}
   By Eqs. (\ref{WaveFunct1}) and (\ref{InitiState2}) the wave function $\Psi^{(0)}(x,t)$ in the  region $\caL^{(0)}$ at time $t$ is expressed as 
\begin{eqnarray}
   \Psi^{(0)}(x,t) = \int_{0}^{L} dx'\; G^{(0,0)}(x,x';t) \Psi^{(0)}(x',0) 
\label{WaveFunct2}
\end{eqnarray}
by using the functions $G^{(0,0)}(x,x';t)$ and $\Psi^{(0)}(x',0)$ for this finite region only. 

   As shown in Appendix \ref{GreenFunctDeriv}, the Green function $G^{(0,0)}(x,x';t)$ is expressed as 
\begin{eqnarray}
   G^{(0,0)}(x,x';t) = \int_{-\infty}^{+\infty} dk \; F(x,x';k) e^{-i \tau_{t} k^{2}} .
\label{GreenFunct2}
\end{eqnarray}
with $\tau_{t}$ defined by 
\begin{eqnarray}
   \tau_{t} &\equiv& \frac{\hbar t}{2m} .
      \label{FunctTau1}
\end{eqnarray}
Here, $F(x,x';k)$ is defined by 
%
\begin{widetext}
\begin{eqnarray}
   F(x,x';k) &\equiv&  
      \frac{2N}{\pi C(k)}\Biggl\{\cos[k(x-x')]
      +\frac{\left(1-\frac{\mu^{2}}{k^{2}}\right)
      \cos[k(x+x'-2L)]+2\frac{\mu}{k}\sin[k(x+x'-2L)]}{1+\frac{\mu^{2}}{k^{2}}}\Biggr\}
      \label{FunctF1}
\end{eqnarray}
with
\begin{eqnarray}
   C(k) \equiv \left|N+1+i\frac{\lambda}{k}+\left(N-1+i\frac{\lambda}{k}\right)
      \frac{1+i\frac{\mu}{k}}{1-i\frac{\mu}{k}}e^{2ikL}\right|^{2} .
\end{eqnarray}
\end{widetext}
It may be noted that the integral region of the wave number $k$ in the right-hand side of Eq. (\ref{GreenFunct2}) is $(-\infty,+\infty)$, different from in the right-hand sides of Eqs. (\ref{GreenFunct1}). 
 
   Noting that the function (\ref{FunctF1}) is an even function of $k$, we expand it as 
\begin{eqnarray}
   F(x,x';k) = \sum_{\nu = 0}^{+\infty} B_{\nu}(x,x') k^{2\nu} 
\label{FunctF2}
\end{eqnarray}
with respect to $k^{2}$, where $B_{\nu}(x,x')$ is a function of $x$ and $x'$ and is independent of $k$. 
   Then, for $t>0$ the $k$-integral in Eq. (\ref{GreenFunct2}) can be carried out, and we obtain 
\begin{eqnarray}
   G^{(0,0)}(x,x';t) &=& \sum_{\nu = 0}^{+\infty} B_{\nu}(x,x') \frac{(2\nu)!}{2^{2\nu}\nu!}
   \frac{1-i}{i^{\nu}}
   \sqrt{\frac{\pi}{2\tau_{t}^{2\nu+1}}} ,
   \nonumber\\
\label{GreenFunct3}
\end{eqnarray}
noting $0!\equiv 1$.
   Here, we used the integral formula
\begin{eqnarray}
   \int_{-\infty}^{+\infty}dk \; k^{2\nu} e^{-i\tau_{t}k^{2}} = \frac{(2\nu)!}{2^{2\nu}\nu!}\frac{1-i}{i^{\nu}}\sqrt{\frac{\pi}{2\tau_{t}^{2\nu+1}}}
\end{eqnarray}
for $\nu = 0,1,\cdots$, which is derived from the $\tau_{t}$-derivative of the equation $\int_{-\infty}^{+\infty}dk \; e^{-i\tau_{t}k^{2}} = (1-i)\sqrt{\pi/(2\tau_{t})}$ for $t>0$. 
   Eq. (\ref{GreenFunct3}) is an asymptotic expansion of the Green function $G^{(0,0)}(x,x';t)$ with respect to $1/t^{\nu+1/2}$, $\nu = 0,1,\cdots$.

   Under the initial condition (\ref{InitiState2}) we describe escape behaviors of a particle from the finite region $\caL^{(0)}$. 
   As an example of quantities to characterize such escape behaviors, we consider the quantity $P(t)$ defined by 
\begin{eqnarray}
   P(t) &\equiv& \frac{\int_{0}^{L} dx \; \rho^{(0)}(x,t)}{\int_{0}^{L} dx \; \rho^{(0)}(x,0)} .
\label{SurviProba1}
\end{eqnarray}
   This is the ratio for a particle to survive in the finite region $\caL^{(0)}$ at time $t$ in comparison with that at the initial time $t=0$ \cite{Memo2}, and we call this probability $P(t)$ the survival probability in this paper \cite{Memo4}. 
   As another quantity to characterize particle escape behaviors we also consider the local velocity for the particle to escape from the finite region $\caL^{(0)}$, which is defined by 
\begin{eqnarray}
   V(t) &\equiv&  
      \frac{ \lim_{x\rightarrow +0} \sum_{n=1}^{N} \rho^{(n)}(x,t)v^{(n)}(x,t)}{\rho(0,t)} 
      \nonumber \\
   &=& - \lim_{x\rightarrow +0} v^{(0)}(x,t) 
\label{LocalVeloc2}
\end{eqnarray}
with the probability density $\rho(0,t)\equiv |\Psi(0,t)|^{2}$ at the junction $O$. 
Here, we used Eq. (\ref{BoundMomen1}) to derive the second equation in the right-hand side of Eq. (\ref{LocalVeloc2}) from its first equation. 
   We call this velocity $V(t)$ the ``escape velocity'' in this paper. 
   The survival probability (\ref{SurviProba1}) and the escape velocity (\ref{LocalVeloc2}) can be calculated by the wave function (\ref{WaveFunct2}) only  in the finite region $\caL^{(0)}$ via the Green function (\ref{GreenFunct3}), without information on the particle in the semi-infinite region $\caL^{(n)}$, $n=1,2,\cdots,N$. 

   In the following subsection, we consider properties of the survival probability $P(t)$ and the escape velocity $V(t)$ mainly in the fixed boundary case $\mu\rightarrow+\infty$, while we give some analytical arguments of $P(t)$ and $V(t)$ in the open boundary case $\mu=0$ in Appendix \ref{OpenBoundaryCase}.

\subsection{Asymptotic properties of particle escapes} 

   In the fixed boundary case $\mu\rightarrow+\infty$, the function (\ref{FunctF1}) is represented as 
\begin{eqnarray}
   F(x,x';k) = \frac{4N}{\pi}\frac{\sin[k(x-L)]\sin[k(x'-L)]}{C(k)}
\label{FunctF3}
\end{eqnarray}
where $C(k)$ is given by 
%
\begin{widetext}
\begin{eqnarray}
   C(k) &=& \left[N+1-(N-1)\cos(2kL)+\lambda\frac{\sin(2kL)}{k}\right]^{2} 
   + \left[(N-1)\sin(2kL)-\lambda\frac{1-\cos(2kL)}{k}\right]^{2} .
\label{FunctC1}
\end{eqnarray}
\end{widetext}
The expansion of the function (\ref{FunctF3}) with respect to $k^{2}$ as in Eq. (\ref{FunctF2}) is different between the cases of $\lambda L \neq -1$ and  $\lambda L = -1$, so we consider these two cases separately below.

\subsubsection{Fixed boundary case with $\lambda L \neq -1$}

\begin{widetext}
   First, we consider the case of $\lambda L \neq -1$.
   For the function (\ref{FunctF3}), the quantity $B_{\nu}(x,x')$ as the coefficients of the function $F(x,x';k)$ with respect to $k^{2\nu}$ in Eq. (\ref{FunctF2}) is given by
\begin{eqnarray}
   B_{0}(x,x') &=& 0 ,
      \\
   B_{1}(x,x') &=& \frac{N(x-L)(x'-L)}{\pi(1+\lambda L)^{2}} ,
      \\
   B_{2}(x,x') &=& - \frac{N(x-L)(x'-L)}{6\pi(1+\lambda L)^{2}}
       \left[(x-L)^{2}+(x'-L)^{2}
       +2L^{2}\frac{3(N^{2}-1)-\lambda L(\lambda L+4)}{(1+\lambda L)^{2}}\right] ,
\end{eqnarray}
$\cdots$, concretely. 
   By inserting these coefficients  $B_{\nu}(x,x')$, $\nu = 0,1,\cdots$ into the formula (\ref{GreenFunct3}) and by using Eq. (\ref{WaveFunct2}) we obtain the wave function $\Psi^{(0)}(x,t)$ in $\caL^{(0)}$ for $t>0$ as 
\begin{eqnarray}
   \Psi^{(0)}(x,t) &=& 
    - \frac{N(x-L)(1+i)\Theta_{1}}{2\sqrt{2\pi}(1+\lambda L)^{2}} 
    \frac{1}{\tau_{t}^{3/2}}
      \nonumber \\
   &&\spaEq
     + \frac{N(x-L)(1-i)}{8\sqrt{2\pi}(1+\lambda L)^{2}}
       \frac{1}{\tau_{t}^{5/2}}
       \left[(x-L)^{2}\Theta_{1}+\Theta_{3}
       +2L^{2}\frac{3(N^{2}-1)-\lambda L(\lambda L+4)}{(1+\lambda L)^{2}}\Theta_{1}\right] 
       +\cdots
       \;\;\;       
\label{WaveFunct3}
\end{eqnarray}
\end{widetext}
where $\Theta_{\nu}$ is defined by
\begin{eqnarray}
   \Theta_{\nu}  \equiv \int_{0}^{L} dx \; (x-L)^{\nu}\Psi^{(0)}(x,0) .
\label{FunctTheta1}
\end{eqnarray}
Eq. (\ref{WaveFunct3}) is the wave function for the fixed boundary case $\mu\rightarrow+\infty$ with $\lambda L \neq -1$ as an expansion of $1/t^{\nu + 1/2}$, $\nu = 1,2,\cdots$. 

   By Eqs. (\ref{ProbaDensi1}), (\ref{FunctTau1}) and (\ref{WaveFunct3}) the survival probability (\ref{SurviProba1}) in the long time limit is expressed asymptotically as  
\begin{eqnarray}
   P(t)  \;\overset{t\rightarrow+\infty}{\sim}\; 
   \frac{2N^{2}\Xi_{1}}{3\pi(1+\lambda L)^{4}} \left(\frac{mL}{\hbar t}\right)^{3} 
\label{SurviAsym1}
\end{eqnarray}
with the constant $\Xi_{1}\equiv \Theta_{1}^{2}/\int_{0}^{L} dx \; \rho^{(0)}(x,0)$ determined by the initial wave function $\Psi^{(0)}(x,0)$ only. 
   Eq. (\ref{SurviAsym1}) means that the survival probability $P(t)$ decays in power $\sim t^{-3}$ asymptotically in time $t$. 
   Moreover, by Eq. (\ref{SurviAsym1}) we obtain 
\begin{eqnarray}
   \lim_{t\rightarrow +\infty} \frac{P(t)}{P(t)|_{\lambda=0}} &=& \frac{1}{(1+\lambda L)^{4}}, 
      \label{SurviPrope1}\\
   \lim_{t\rightarrow +\infty} \frac{P(t)}{P(t)|_{N=1}} &=& N^{2} .
      \label{SurviPrope2}
\end{eqnarray}
as far as the initial wave function  $\Psi^{(0)}(x,0)$ is independent of $\lambda$ and $N$. 
   Eq. (\ref{SurviPrope1}) shows that the probability for a particle to stays in the finite region $\caL^{(0)}$ becomes lower for a larger amplitude (\ref{AmpliPoten1}) of potential barrier at the junction $O$ in the long time limit, and this (rather counter-intuitive) result, as well as the asymptotic power decay $\sim t^{-3}$ of $P(t)$, has already been shown for the one-lead case $N=1$ in Ref. \cite{GMV07}. 
   On the other hand, Eq. (\ref{SurviPrope2}) means that the probability for a particle to stays in the finite region $\caL^{(0)}$ for multi semi-infinite leads becomes $N^{2}$ times higher than that for a single lead. 

   The first non-zero contribution of the survival probability $P(t)$ after a long time comes from the first term of the right-hand side of Eq. (\ref{WaveFunct3}). 
   In contrast, the escape velocity $V(t)$ is an example of quantities in which the first non-zero contribution after a long time comes from the second term of the right-hand side of Eq. (\ref{WaveFunct3}). 
   Actually, by inserting the right-hand side of Eq. (\ref{WaveFunct3}) up to its second term into Eq.   (\ref{LocalVeloc1}), and using Eq.  (\ref{FunctTau1}) we obtain 
\begin{eqnarray}
   v^{(0)}(x,t) \;\overset{t\rightarrow+\infty}{\sim}\; -\frac{L-x}{t}
\label{LocalVeloc3}
\end{eqnarray}
asymptotically in time. 
   Therefore, after a long time the escape velocity (\ref{LocalVeloc2}) is represented as  
\begin{eqnarray}
   V(t) \;\overset{t\rightarrow+\infty}{\sim}\; \frac{L}{t} ,
\label{AsympEscapVeloc1}
\end{eqnarray}
meaning that asymptotically in time, the escape velocity $V(t)$ decays in power $\sim t^{-1}$ and is independent of the number $N$ of semi-infinite leads, the constant $\lambda$, and the initial wave function $\Psi^{(0)}(x,0)$, etc. 
   The velocity (\ref{AsympEscapVeloc1}) can be regarded as a constant velocity by which a classical mechanical particle in an ideal wire remains inside a finite region with the length $L$ within the time interval $t$ without an escape.

\subsubsection{Fixed boundary case with $\lambda L = -1$}

   Next, we consider the case of $\lambda L = -1$, in which the quantities $B_{\nu}(x,x')$, $\nu=0,1,\cdots$ are represented as
%
\begin{widetext}
\begin{eqnarray}
   B_{0}(x,x') &=& \frac{(x-L)(x'-L)}{\pi N L^{2}} ,
      \\
   B_{1}(x,x') &=& -\frac{(x-L)(x'-L)}{6\pi N L^{2}}\left[(x-L)^{2}+(x'-L)^{2}
      -\frac{2L^{2}}{3}\frac{3N^{2}-1}{N^{2}}\right] ,
\end{eqnarray}
$\cdots$, concretely, by Eqs. (\ref{FunctF2}), (\ref{FunctF3}) and (\ref{FunctC1}). 
   By inserting these coefficients $B_{\nu}(x,x')$, $\nu = 0,1,\cdots$ of the function $F(x,x';t)$ with respect to $k^{2}$ into Eq. (\ref{GreenFunct3}) the wave function (\ref{WaveFunct2}) in the region $\caL^{(0)}$ for $t>0$ is represented as 
\begin{eqnarray}
   \Psi^{(0)}(x,t) &=& 
       \frac{(x-L)(1-i)\Theta_{1}}{\sqrt{2\pi} N L^{2}} \frac{1}{\tau_{t}^{1/2}}
       +\frac{(x-L)(1+i)}{12 \sqrt{2\pi} N L^{2}}
       \left[ (x-L)^{2} \Theta_{1} + \Theta_{3}-\frac{2L^{2}}{3}\frac{N^{2}-1}{N^{2}} 
       \Theta_{1}\right] 
       \frac{1}{\tau_{t}^{3/2}}
       +\cdots
\label{WaveFunct5}
\end{eqnarray}
\end{widetext}
as an expansion with respect to $1/t^{\nu+1/2}$, $\nu = 0,1,\cdots$. 

   From Eqs. (\ref{ProbaDensi1}), (\ref{FunctTau1}), (\ref{SurviProba1}) and (\ref{WaveFunct5}) we derive the survival probability 
\begin{eqnarray}
   P(t)  \;\overset{t\rightarrow+\infty}{\sim}\; 
   \frac{2m\Xi_{1}}{3\pi\hbar N^{2}Lt}
\label{SurviAsym3}
\end{eqnarray}
asymptotically in time. 
   It is important to note that the survival probability (\ref{SurviAsym3}) for $\lambda L = -1$ decays in power $\sim t^{-1}$ asymptotically in time, different from the asymptotic decay power $\sim t^{-3}$ for $\lambda L \neq -1$ as shown in Eq. (\ref{SurviAsym1}). 
   By the asymptotic form of the survival probability (\ref{SurviAsym3}) we obtain 
\begin{eqnarray}
   \lim_{t\rightarrow +\infty} \frac{P(t)}{P(t)|_{N=1}} &=& \frac{1}{N^{2}}
      \label{SurviPrope3}
\end{eqnarray}
for the $N$-independent initial wave function  $\Psi^{(0)}(x,0)$. 
Eq. (\ref{SurviPrope3}) means that by connecting more number $N$ of the leads to the finite region $\caL^{(0)}$, the probability of a particle to stay in the region $\caL^{(0)}$ for  $\lambda L = -1$ becomes lower after a long time, opposite to Eq. (\ref{SurviPrope2}) for $\lambda L \neq -1$. 
   
   From Eqs. (\ref{LocalVeloc1}), (\ref{FunctTau1}) and (\ref{WaveFunct5}) we derive the local velocity $v^{(0)}(x,t)$ in $\caL^{(0)}$ as 
\begin{eqnarray}
   v^{(0)}(x,t) \;\overset{t\rightarrow+\infty}{\sim}\; -\frac{L-x}{3t}
\label{LocalVeloc3b}
\end{eqnarray}
asymptotically in time. Therefore, we obtain the escape velocity (\ref{LocalVeloc2}) as 
\begin{eqnarray}
   V(t) \;\overset{t\rightarrow+\infty}{\sim}\; \frac{L}{3t} 
\label{AsympEscapVeloc3}
\end{eqnarray}
after a long time. Eq. (\ref{AsympEscapVeloc3}) shows that the escape velocity $V(t)$ for  $\lambda L = -1$ decays asymptotically in the same power $\sim t^{-1}$ as in Eq. (\ref{AsympEscapVeloc1}) for $\lambda L \neq -1$, although its prefactor $L/3$ is one third of that in the case of $\lambda L \neq -1$.

\subsection{Finite time properties of particle escapes}
\label{FiniteTimeProperties}

   In this subsection, we consider a finite time behavior of particle escapes in the fixed boundary case $\mu\rightarrow +\infty$ by calculating the wave function $\Psi^{(0)}(x,t)$ numerically. 
   In order to calculate the wave function $\Psi^{(0)}(x,t)$ concretely for the fixed boundary case, we specify the initial wave function in the finite region $\caL^{(0)}$ as 
\begin{eqnarray}
   \Psi^{(0)}(x,0) = \sqrt{\frac{2}{L}}\sin\left(\frac{\sigma\pi}{L} x \right) ,
\label{InitiState3}
\end{eqnarray}
which is the $\sigma$-th eigenstate of a particle confined in the finite region $\caL^{(0)}$ without semi-infinite leads ($\sigma = 1,2,\cdots$) \cite{Memo6}.
   Under the initial condition (\ref{InitiState2}) and (\ref{InitiState3}), by Eqs. (\ref{WaveFunct2}), (\ref{GreenFunct2}) and (\ref{FunctF3}) the wave function $\Psi^{(0)}(x,t)$ in the finite region $\caL^{(0)}$ at time $t$ is represented as
%
\begin{widetext}
\begin{eqnarray}
   \Psi^{(0)}(x,t) = 8\sigma N\sqrt{2L}\int_{0}^{+\infty} dk\; 
   \frac{e^{-i\tau_{t}k^{2}}\sin(kL)\sin[k(x-L)]}{C(k)\left[(kL)^{2}-(\sigma \pi)^{2}\right]}
\label{WaveFunct6}
\end{eqnarray}
\end{widetext}
with $C(k)$ given by Eq. (\ref{FunctC1}). 
   By carrying out the integral (\ref{WaveFunct6}) with respect to $k$ numerically, we calculate the serval probability $P(t)$ and the escape velocity $V(t)$ for a finite time $t$. 
   In this subsection we chose the parameter values as $L=1$, $\sigma =1$, $m=1$ and $\hbar =1$.

\subsubsection{Fixed boundary case with $\lambda L \neq -1$} 

\begin{figure}[!t]
\vspfigA
\begin{center}
\includegraphics[width=\widthfig]{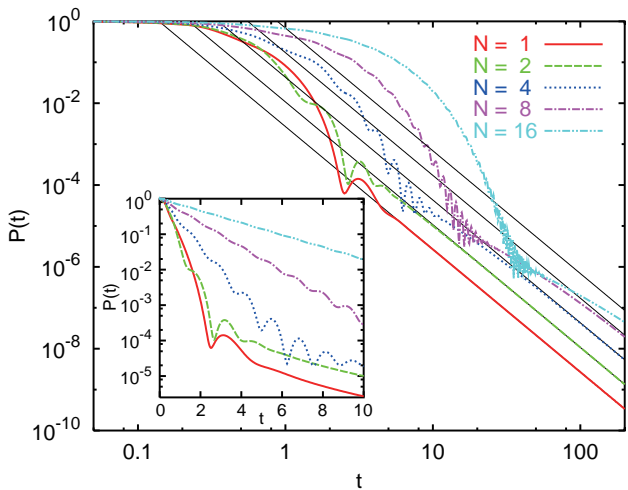}
\vspfigB
\caption{
   Survival probability $P(t)$ as a function of $t$ for the fixed boundary case with $\lambda L\neq -1$ for $N=1$ (the solid line), $2$ (the broken line), $4$ (the dotted line), $8$ (the dash-dotted line) and $16$ (the dash-double-dotted line). 
   The main figure is log-log plots of $P(t)$ with the corresponding asymptotic power decays shown in thin straight lines, and the inset is their linear-log plots for a short time.  
}
\label{Fig2aFixASurProb}
\end{center}
\vspfigC
\end{figure}  
%
\begin{figure}[!t]
\vspfigA
\begin{center}
\includegraphics[width=\widthfig]{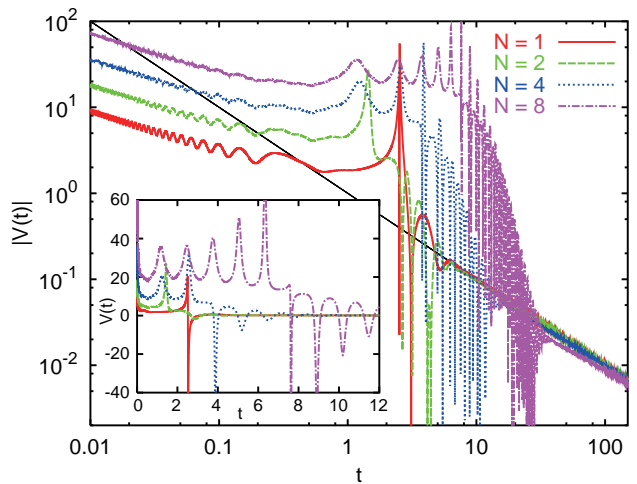}
\vspfigB
\caption{
   Absolute value $|V(t)|$ of the escape velocity as a function of $t$ for the fixed boundary case with $\lambda L\neq -1$ for $N=1$ (the solid line), $2$ (the broken line), $4$ (the dotted line) and $8$ (the dash-dotted line). 
   The main figure is log-log plots of $|V(t)|$ with thin straight lines to show the asymptotic  power decays, and the inset is linear-linear plots of the escape velocity $V(t)$ itself for a short time. 
}
\label{Fig2bFixAEscVel}
\end{center}
\vspfigC
\end{figure}  
%
   Figure \ref{Fig2aFixASurProb} is the survival probability $P(t)$ as a function of time $t$ for the fixed boundary case $\mu\rightarrow +\infty$ with $\lambda L \neq -1$ under the initial wave function (\ref{InitiState3}) for different numbers $N=1$ (the solid line), $2$ (the broken line), $4$ (the dotted line), $8$ (the dash-dotted line) and $16$ (the dash-double-dotted line)  of semi-infinite leads. 
   Here, we used the parameter value $\lambda =1$, and the thin straight lines in this figure show the asymptotic power decay (\ref{SurviAsym1}) of the survival probability for each value of $N$. 

   The numerical results in the main figure of Fig. \ref{Fig2aFixASurProb} as log-log plots of $P(t)$  show that the survival probability $P(t)$ approaches to the power decay (\ref{SurviAsym1}) asymptotically in time from an earlier time for a smaller number $N$ of leads. 
   We can also see in the inset of Fig. \ref{Fig2aFixASurProb} as linear-log plots of $P(t)$ that for a short time the survival probability $P(t)$ decays exponentially $\sim \exp(-\alpha t)$ in time with a positive constant $\alpha$, as shown especially in the cases of $N=4$, $8$ and $16$. 
   The time period for such an exponential decay of $P(t)$ is longer for more number $N$ of leads, and its decay rate $\alpha$ is smaller for more number $N$ of leads. 
   One may notice that the survival probability does not decrease monotonously in time and shows a time-oscillation between its exponential decay region and the power decay region.  
   As a tendency, the survival probability $P(t)$ is higher for more number $N$ of leads, although there are temporal exceptions for it by its time-oscillatory behavior.

   Figure \ref{Fig2bFixAEscVel} is plots of the absolute value  $|V(t)|$ of the escape velocity as the main figure, as well as the escape velocity $V(t)$ itself as the inset,  as a function of time $t$ for the fixed boundary case under the initial wave function (\ref{InitiState3}) for the different lead number $N=1$ (the solid line), $2$ (the broken line), $4$ (the dotted line) and $8$ (the dash-dotted line). 
   Here, we used the parameter value $\lambda =1$, and the thin straight line is the asymptotic form (\ref{AsympEscapVeloc1}) of the escape velocity. 

   It is shown in the inset of Fig. \ref{Fig2bFixAEscVel} that in a short time the escape velocity $V(t)$ oscillates in time, rather than a simple decay, and can take even a negative value sometimes. 
   Such a time-oscillatory behavior of $V(t)$ continues for a longer time for more number $N$ of leads, although its time oscillating period seems to be almost independent of $N$. 
   For a short time, as a tendency the magnitude of the escape velocity $V(t)$ becomes larger for more number $N$ of leads. 
   For some values of $N$, such as for $N=1$ in the inset of Fig. \ref{Fig2bFixAEscVel}, a very rapid (but not abrupt) change of the escape velocity $V(t)$ from a positive value to the first negative value as a function of $t$ occurs, when the value of $|\Psi^{(n)}(+0,t)|$ is non-zero but very small at the time $t$ satisfying the condition $\partial \Psi^{(n)}(x,t)/\partial x|_{x=+0} =0$. 
   (Note that time-oscillations of $V(t)$ around the time $t\approx 10$ in the main figure of Fig. \ref{Fig2bFixAEscVel} look like to be cut off in a middle occasionally, because there are not enough numbers of calculation points, but they actually reach the value of zero by crossing the line $V=0$.) 
   After such a time-oscillation, the escape velocity $V(t)$ converges rapidly to its asymptotic form (\ref{AsympEscapVeloc1}), which is independent of $N$.

\subsubsection{Fixed boundary case with $\lambda L = -1$} 

\begin{figure}[!t]
\vspfigA
\begin{center}
\includegraphics[width=\widthfig]{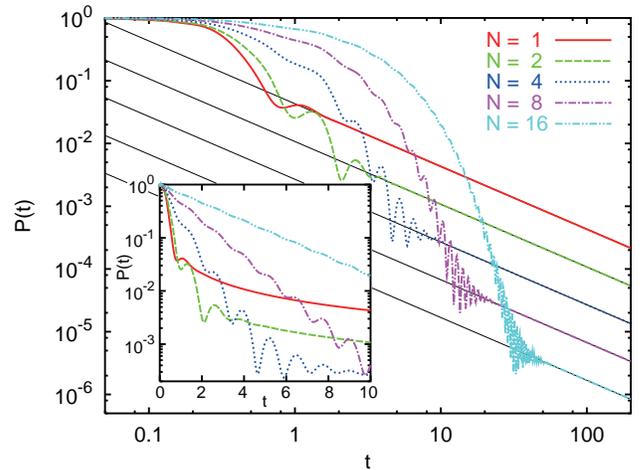}
\vspfigB
\caption{
   Survival probability $P(t)$ as a function of $t$ for the fixed boundary case with $\lambda L= -1$ for $N=1$ (the solid line), $2$ (the broken line), $4$ (the dotted line), $8$ (the dash-dotted line) and $16$ (the dash-double-dotted line). 
   The main figure is log-log plots of $P(t)$ with the corresponding asymptotic power decays shown in thin straight lines, and the inset is their linear-log plots for a short time.  
}
\label{Fig2cFixBSurProb}
\end{center}
\vspfigC
\end{figure}  
%
\begin{figure}[!t]
\vspfigA
\begin{center}
\includegraphics[width=\widthfig]{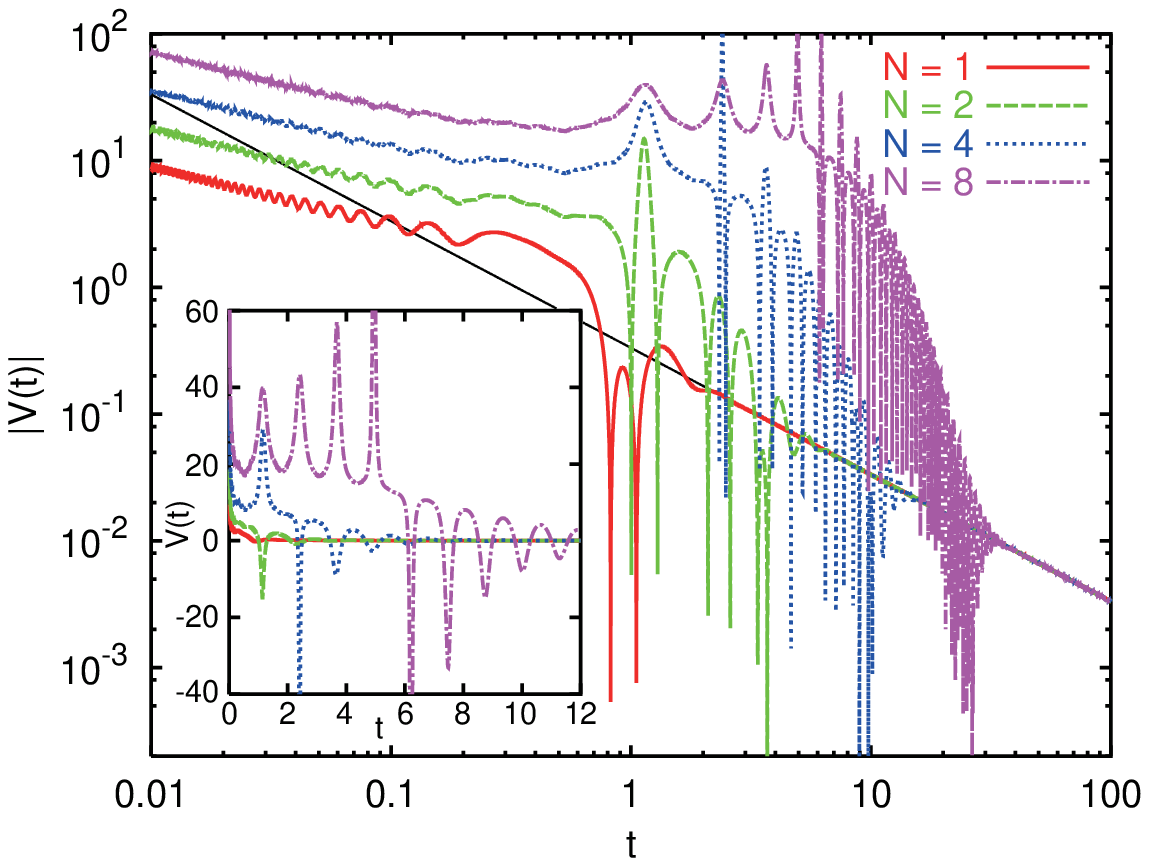}
\vspfigB
\caption{
   Absolute value $|V(t)|$ of the escape velocity as a function of time $t$ for the fixed boundary case with $\lambda L = -1$ for $N=1$ (the solid line), $2$ (the broken line), $4$ (the dotted line) and $8$ (the dash-dotted line). 
   The main figure is log-log plots of $|V(t)|$ with a straight line for the asymptotic power decay, and the inset is liner-linear plots of the escape velocity $V(t)$ itself for a short time.  
}
\label{Fig2dFixBEscVel}
\end{center}
\vspfigC
\end{figure}  
%
   Now, we consider finite time properties of particle escapes for $\lambda L = -1$ (so $\lambda =-1$ because of $L=1$) in the fixed boundary case $\mu\rightarrow +\infty$. 
   Figure \ref{Fig2cFixBSurProb} is the survival probability $P(t)$ as a function of time $t$ for $N=1$ (the solid line), $2$ (the broken line), $4$ (the dotted line), $8$ (the dash-dotted line) and $16$ (the dash-double-dotted line). 
   The thin straight lines in this figure show the asymptotic form (\ref{SurviAsym3}) of $P(t)$ for each value of $N$. 

   As shown in the inset of Fig. \ref{Fig2cFixBSurProb} as linear-log plots of $P(t)$, especially for large $N = 8$ and $16$, in a short time region the survival probability $P(t)$ decays exponentially in time. 
   Such an exponential decay continues for a longer time for more number $N$ of attached leads, and its decay rate is smaller for more number $N$ of leads. 
   After the exponential decay, the survival probability $P(t)$ shows a time-oscillatory behavior, then converges to its asymptotic power decay form (\ref{SurviAsym3}), as shown in the main figure of Fig. \ref{Fig2cFixBSurProb} as log-log plots of $P(t)$. 
   Contrast to the case of  $\lambda L \neq -1$, the survival probability $P(t)$ becomes lower  for more number of $N$ in this asymptotic power decay region, while it is opposite in the exponential time-decay region. 
   
   Figure \ref{Fig2dFixBEscVel} is the absolute value  $|V(t)|$ of the escape velocity as the main figure, as well as the escape velocity $V(t)$ itself as the inset,  as a function of time $t$ for the fixed boundary condition with $\lambda L = -1$ in the case of  $N=1$ (the solid line), $2$ (the broken line), $4$ (the dotted line) and $8$ (the dash-dotted line). 
   The thin straight line in this figure is the asymptotic form (\ref{AsympEscapVeloc3}) of the escape velocity $V(t)$. 
   The escape velocity $V(t)$ itself is also shown for a short time in the inset of this figure. 
   
   As shown in the inset of Fig. \ref{Fig2dFixBEscVel}, the escape velocity $V(t)$ oscillates in time, first as a time-oscillation with positive values then as that with positive and negative values. 
   Such an oscillatory time region continues for a longer time for more number $N$ of leads, although its time oscillating period seems to be almost independent of $N$, and as a tendency the absolute value $|V(t)|$ of the escape velocity is larger for more number $N$ of leads in this time region. 
   (In the main figure of Fig. \ref{Fig2dFixBEscVel}, the time-oscillations of $V(t)$ look to be cut off in a middle, because of little number of calculation points. In the actual graphs the value of $|V(t)|$ go to zero in a time-oscillatory region with positive and negative values.) 
   After such a time-oscillation the escape velocity $V(t)$ converges rapidly to its asymptotic power decay form (\ref{AsympEscapVeloc3}), which is independent of $N$.

\section{Conclusion and remarks} 
\label{ConclusionRemarks}

   In this paper, we have discussed particle escapes in an open quantum network system by a scattering theoretical approach. 
   As a concrete example with a current splitter as a feature of network systems, we considered particle escapes from an end of a finite one-dimensional wire to $N$ number of semi-infinite one-dimensional leads. 
   Properties of particle escapes in such a quantum network was discussed by using the two kinds of quantities; the one is the probability $P(t)$ for the particle to remain in the finite wire at time $t$, the so-called survival probability, and the other is the velocity $V(t)$ for the particle to leave from the finite region, the so-called escape velocity. 
   Here, the escape velocity $V(t)$ is introduced from the probability current, based on the equation of continuity for the particle position probability density. 
   With the fixed boundary condition at an end of the finite lead, for the potential barrier amplitude $\Lambda \neq -\hbar^{2}/(2mL)$ the survival probability $P(t)$ depends on the number $N$ of attached semi-infinite leads as $\lim_{t\rightarrow +\infty} P(t)/P(t)|_{N=1} = N^{2}$ and decays in power $\sim 1/t^{3}$ asymptotically in time. 
   In contrast, for the potential barrier amplitude $\Lambda = -\hbar^{2}/(2mL)$ the survival probability satisfies the relation $\lim_{t\rightarrow +\infty} P(t)/P(t)|_{N=1} = 1/N^{2}$, and it decays in power $\sim 1/t$ after a long time. 
   On the other hand, the escape velocity $V(t)$ decays like $C L/t$ asymptotically in time with the constant $C$ which is independent of the number $N$ of leads and the initial wave function $\Psi^{(0)}(x,0)$, etc. 
   It was also shown that for a finite time the survival probability $P(t)$ decays exponentially in time for a longer time with a smaller decay rate for more number $N$ of attached leads, and shows a time-oscillatory behavior between the exponential decay time region and the power decay time region. 
   The escape velocity $V(t)$ show a time-oscillatory behavior for a finite time, and as a tendency its value is higher for more number $N$ of attached leads with a larger amplitude of time-oscillations.

   We described the dynamics of an escaping particle by a quantum scattering theoretical approach. 
   It may be noted that although the time-dependent wave function of an escaping particle is expanded by the scattering states in the open network system it is nonzero only in a finite region for a finite time and is normalizable, different from stationary quantum scattering states caused by incident plain waves from infinitely far spatial regions.   
   In order to construct concretely the scattering states in the open network system, it is essential to specify the boundary conditions at the junction of a finite wire and multiple leads and at another end of the finite wire. 
   We specified these boundary conditions based on the probability current given from the equation of continuity for the particle position probability density. 
   We imposed the conservation of this current and the continuities of the particle wave function at the junction, so that the scattering matrix at the junction is unitary and the scattering states satisfies the orthogonal relation automatically. 
   It is important to note that the condition of no net current at an end of the finite wire leads to a group of boundary conditions specified by a parameter $\mu$. 
   In this paper we mainly considered the case of $\mu\rightarrow+\infty$, i.e. of the fixed boundary condition. 
   However, by using the case of $\mu = 0$, i.e. of the open boundary condition, we can also discuss particle escapes from a finite one-dimensional wire whose \textit{both} ends are connected with multiple one-dimensional semi-infinite leads. 
   In this case, for the initial state which is anti-symmetric with the reflection at the center of the finite region, we get the same results of the survival probability $P(t)$ and the escape velocity $V(t)$ as cases of the fixed boundary condition. 
   In contrast, for the initial state which is symmetric with the reflection at the center of the finite region, by using the open boundary condition $\mu =0$ we obtain  $\lim_{t\rightarrow +\infty}P(t)/[P(t)|_{N=1}]=N^{2}$ and  $V(t) \;\overset{t\rightarrow+\infty}{\sim}\; 3L/t$ for $\Lambda \neq 0$, and $\lim_{t\rightarrow +\infty}P(t)/[P(t)|_{N=1}]=1/N^{2}$ and  $V(t) \;\overset{t\rightarrow+\infty}{\sim}\; L/t$   for  $\Lambda = 0$, in which different behaviors of $P(t)$ and $V(t)$ occur at a different value of $\Lambda$ from that in the anti-symmetric initial state. 
   As a remark it may be interesting if we could clarify the physical meanings of particle escapes in the case of other values of the parameter $\mu$, i.e. a nonzero and finite values of $\mu$. 
   We also note that the current used to specify these boundary conditions is a current of the probability density of the particle position. In this sense, the escape velocity $V(t)$ based on this probability current at the junction is not the particle velocity itself. 
   The uncertain principle of quantum mechanics forbids to specify the particle velocity at a specific position of particle.

\begin{figure}[!t]
\vspfigA
\begin{center}
\includegraphics[width=\widthfig]{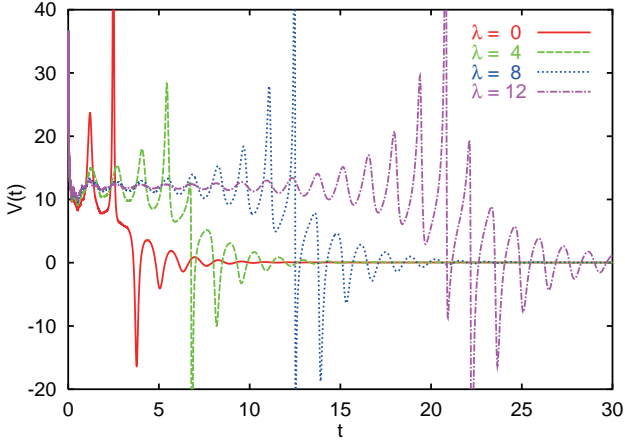}
\vspfigB
\caption{
   Escape velocity $V(t)$ as a function of $t$ for the fixed boundary case with $\lambda=0$ (the solid line), $4$ (the broken line), $8$ (the dotted line) and $12$ (the dash-dotted line). 
}
\label{Fig3aFixAEscVel}
\end{center}
\vspfigC
\end{figure} 
%
   As finite time properties of quantum escapes, in this paper we discussed mainly dependences of the survival probability $P(t)$ and the escape velocity $V(t)$ on the number $N$ of attached leads, but their dependences on other parameters also express some important escape properties. 
   For example, the survival probability $P(t)$ decays exponentially for a finite time with a smaller decay rate for a larger value of the parameter $\lambda$ proportional to the potential barrier amplitude, as known already for the case of $N=1$ \cite{OK91,TS11}. 
   As another example, we show Fig. \ref{Fig3aFixAEscVel} for the escape velocity $V(t)$ as a function of time $t$ for the fixed boundary case $\mu\rightarrow +\infty$ with $\lambda=0,4,8$ and $12$. 
   Here, we chose the other parameter values as $L=1$, $N=4$, $\sigma =1$, $m=1$ and $\hbar =1$. 
   This figure shows that the escape velocity $V(t)$ oscillates in time around a constant value $V_{0}$ for a short time, and the value $V_{0}$ is almost independent of $\lambda$ but the time period appearing such time-oscillations around the value $V_{0}$ becomes longer for a larger value of the parameter $\lambda$. 
   Detailed dependences of escape behaviors in quantum networks via multiple leads on other parameters, such as $\lambda$, $\mu$ and $\sigma$, etc., will be discussed elsewhere.

   In this paper we discussed briefly exponential decays of the survival probability for a finite time, appearing especially for a large number $N$ of attached leads as well as for a large value of the potential barrier amplitude $\Lambda$. These exponential decays would be related to the poles of the Green function or the scattering matrix \cite{DN02,CM97,GMV07,OK91,BG01}. A detailed analysis of such exponential decays of the survival probability as the scattering resonance in particle escapes via multiple leads remains as an important future problem.

\section*{Acknowledgements}

   On of the authors (T.T) is grateful to T. Okamura for many valuable comments and suggestions on decay behaviors in quantum open systems. 
   He also thanks P. A. Jacquet for discussions on quantum escape problems. 
   This research was supported by the grant sponsor: The "Open Research Project for Physical Science of Biomolecular Systems" funded by the Ministry of Education, Culture, Sports, Science, and Technology of Japan.

\setcounter{section}{0} 
\makeatletter 
   \@addtoreset{equation}{section} 
   \makeatother 
   \def\theequation{\Alph{section}.%
   \arabic{equation}} 
  
\appendix

\section{Scattering matrix}
\label{PropeScattMatri}

   In this appendix we give a derivation of Eq. (\ref{ScattMatri1}) with the scattering matrix (\ref{ScattMatri2}).   
   
   By inserting Eq. (\ref{ScattState1}) into the condition (\ref{BoundPosit1}) we obtain 
\begin{eqnarray}
   a_{(j,k)}^{(0)} + b_{(j,k)}^{(0)} 
   &=&
   a_{(j,k)}^{(1)} + b_{(j,k)}^{(1)} 
   \nonumber \\
   &=&
   \cdots =a_{(j,k)}^{(N)} + b_{(j,k)}^{(N)} =  \Psi(0,t) .
   \;\;\;
\label{BoundPosit2}
\end{eqnarray}
On the other hand, Eqs. (\ref{BoundVeloc1}) and (\ref{ScattState1}) lead to  
\begin{eqnarray}
   \sum_{n=0}^{N}\left[a_{(j,k)}^{(n)} - b_{(j,k)}^{(n)}\right]
   = i\frac{\lambda}{k} \Psi(0,t) .
\label{BoundVeloc2}
\end{eqnarray}
By Eqs. (\ref{BoundPosit2}) and (\ref{BoundVeloc2}) we obtain $\sum_{n=0}^{N}\left[2a_{(j,k)}^{(n)} - \Psi(0,t) \right]
   = i\frac{\lambda}{k} \Psi(0,t) $, i.e.
\begin{eqnarray}
   \Psi(0,t) = \frac{2}{N+1+i\frac{\lambda}{k}}\sum_{n=0}^{N}a_{(j,k)}^{(n)} .
\label{BoundVeloc3}
\end{eqnarray}
From Eqs. (\ref{BoundPosit2}) and (\ref{BoundVeloc3}) we derive
\begin{eqnarray}
   \bfb_{(j,k)} &=& \Psi(0,t) (1 1 \cdots 1)^{T} - \bfa_{(j,k)} 
      \nonumber \\
   &=& S_{k} \bfa_{(j,k)}
\end{eqnarray}
with the scattering matrix (\ref{ScattMatri2}). 
   Therefore, we obtain Eq. (\ref{ScattMatri1}).

\section{Orthogonality of scattering states}
\label{OrthoScattState}

   In this appendix we give a proof of the orthogonality relation (\ref{OrthoScatt1}) for the scattering state $\Phi_{(j,k)}^{(n)}(x)$. 
   
   We note the mathematical identity
\begin{eqnarray}
   \int_{0}^{+\infty} dx\; e^{ikx} 
   &\equiv& \lim_{\epsilon\rightarrow +0} \int_{0}^{+\infty} dx\; e^{ikx-\epsilon x}
   \nonumber \\
   &=& \pi \delta (k) +i \chi(k)
\label{MatheIdent1}
\end{eqnarray}
where $\delta (k)$ is the delta function 
\begin{eqnarray}
   \delta (k)=\lim_{\epsilon\rightarrow +0} \frac{1}{\pi}\frac{\epsilon}{k^{2}+\epsilon^{2}} 
\end{eqnarray}
%
and $\chi(k)$ is defined by 
\begin{eqnarray}
   \chi(k)\equiv \lim_{\epsilon\rightarrow +0} \frac{k}{k^{2}+\epsilon^{2}} .
\label{FunctChi1}
\end{eqnarray}
Using the function (\ref{FunctChi1}) we obtain 
\begin{eqnarray}
   \int_{0}^{L} dx\; e^{ikx} 
   &=& \left(1-e^{ikL}\right) \int_{0}^{+\infty} dx\; e^{ikx} 
      \nonumber \\
   &=& \left(1-e^{ikL}\right) i \chi(k) ,
\label{MatheIdent2}
\end{eqnarray}
where we used Eq. (\ref{MatheIdent1}) and the identity $(1-e^{ikL})\delta (k) =0$. 

   By the scattering state (\ref{ScattState1}) and the inner product (\ref{InneProdu1}) for quantum  states, as well as Eqs. (\ref{ScattMatri1}), (\ref{ScattMatri3}), (\ref{MatheIdent1}) and (\ref{MatheIdent2}),  we obtain 
%
\begin{widetext}
\begin{eqnarray}
      \InPro{\Phi_{(j,k)}}{\Phi_{(j',k')}} 
   &=& \lim_{\epsilon\rightarrow +0}\int_{\epsilon}^{L} dx \; 
      \Phi_{(j,k)}^{(0)}(x)^{*}\Phi_{(j',k')}^{(0)}(x)
      + \lim_{\epsilon\rightarrow +0}\sum_{n=1}^{N}
      \int_{\epsilon}^{+\infty} dx \; \Phi_{(j,k)}^{(n)}(x)^{*}\Phi_{(j',k')}^{(n)}(x) 
      \nonumber \\
   &=& \pi \sum_{n=1}^{N}\left[a_{(j,k)}^{(n)}{}^{*}a_{(j',k')}^{(n)}
      +b_{(j,k)}^{(n)}{}^{*}b_{(j',k')}^{(n)}\right]\delta(k-k')
   + \pi \sum_{n=1}^{N}\left[a_{(j,k)}^{(n)}{}^{*}b_{(j',k')}^{(n)}
      +b_{(j,k)}^{(n)}{}^{*}a_{(j',k')}^{(n)}\right]\delta(k+k')
      \nonumber \\
   &&\spaEq -i \bfa_{(j,k)}^{\dagger}\left[ \left(S_{k}^{\dagger}S_{k'}-I\right)\chi(k-k') 
      + \left(S_{k}^{\dagger}-S_{k'}\right)\chi(k+k') \right]\bfa_{(j',k')}
      \nonumber \\
   &&\spaEq -i b_{(j,k)}^{(0)}{}^{*}\left[ \left(s_{k}^{*}s_{k'}-1\right)\chi(k-k') 
      + \left(s_{k}^{*}-s_{k'}\right)\chi(k+k') \right]b_{(j',k')}^{(0)}e^{-i(k-k')L}
\label{OrthoRemark1}
\end{eqnarray}
\end{widetext}
in which $s_{k}$ is defined by
\begin{eqnarray}
   s_{k}\equiv \frac{1+i\frac{\mu}{k}}{1-i\frac{\mu}{k}} .
\label{WallScatt1}
\end{eqnarray}
   Now, we note 
\begin{eqnarray}
   && \spaNeg 
      \sum_{n=1}^{N}\left[a_{(j,k)}^{(n)}{}^{*}a_{(j',k')}^{(n)}
      +b_{(j,k)}^{(n)}{}^{*}b_{(j',k')}^{(n)}\right]\delta(k-k')
      \nonumber \\
   &&= \left[\bfa_{(j,k)}^{\dagger}\bfa_{(j',k')}
      +\bfb_{(j,k)}^{\dagger}\bfb_{(j',k')}
      \right.\nonumber\\
      &&\spaEq\left.    
      - a_{(j,k)}^{(0)}{}^{*}a_{(j',k')}^{(0)}
      - b_{(j,k)}^{(0)}{}^{*}b_{(j',k')}^{(0)}\right]\delta(k-k')
      \nonumber \\
   &&= 2\left[\bfa_{(j,k)}^{\dagger}\bfa_{(j',k')}
      - a_{(j,k)}^{(0)}{}^{*}a_{(j',k')}^{(0)}\right]\delta(k-k')
      \nonumber \\
   &&= 2\sum_{n=1}^{N}a_{(j,k)}^{(n)}{}^{*}a_{(j',k')}^{(n)}\delta(k-k')
      \nonumber \\
   &&= \frac{1}{\pi}\delta_{jj'}\delta(k-k')
\label{OrthoRemark2}
\end{eqnarray}
by Eqs. (\ref{ScattMatri1}), (\ref{UnitaScatt1}), (\ref{ScattMatri3}), (\ref{ScattCoeff3}) and (\ref{ConstA1}). 
   Moreover, using the function (\ref{FunctChi1}) we note 
\begin{eqnarray}
   \int dk\; \chi(k) X(k) = \hat{\caP} \int dk\; \frac{X(k)}{k}
\label{OrthoRemark3}
\end{eqnarray}
for any function $X(k)$ of $k$, where we introduced the operator $\hat{\caP}$ as that to take the principal integral. 
   In this sense, for the scattering matrix (\ref{ScattMatri2}) we obtain 
\begin{eqnarray}
   && \spaNeg 
       \left(S_{k}^{\dagger}S_{k'}-I\right)\chi(k-k') 
      + \left(S_{k}^{\dagger}-S_{k'}\right)\chi(k+k') 
      \nonumber \\
   &&= \hat{\caP}\left[\left(S_{k}^{\dagger}S_{k'}-I\right)\frac{1}{k-k'} 
      + \left(S_{k}^{\dagger}-S_{k'}\right)\frac{1}{k+k'}\right]
      \nonumber \\
   &&= \caO 
\label{OrthoRemark4}
\end{eqnarray}
with the matrix $\caO$ whose all elements are zero.
Similarly, for the quantity $s_{k}$ defined by Eq. (\ref{WallScatt1}) we obtain 
\begin{eqnarray}
   && \spaNeg 
      \left(s_{k}^{*}s_{k'}-1\right)\chi(k-k') 
      + \left(s_{k}^{*}-s_{k'}\right)\chi(k+k') 
      \nonumber \\
   &&= \hat{\caP}\left[\left(s_{k}^{*}s_{k'}-1\right)\frac{1}{k-k'} 
      + \left(s_{k}^{*}-s_{k'}\right)\frac{1}{k+k'} \right]
      \nonumber \\
   &&= 0 .
\label{OrthoRemark5}
\end{eqnarray}
Inserting Eqs. (\ref{OrthoRemark2}), (\ref{OrthoRemark4}) and (\ref{OrthoRemark5}) into Eq. (\ref{OrthoRemark1}) we obtain
%
\begin{widetext}
\begin{eqnarray}
   \InPro{\Phi_{(j,k)}}{\Phi_{(j',k')}} 
      = \delta_{jj'}\delta(k-k')
      + \pi \sum_{n=1}^{N}\left[a_{(j,k)}^{(n)}{}^{*}b_{(j',k')}^{(n)}
      +b_{(j,k)}^{(n)}{}^{*}a_{(j',k')}^{(n)}\right]\delta(k+k'). 
\label{OrthoRemark6}
\end{eqnarray}
\end{widetext}
Noting that the second term of the right-hand side of Eq. (\ref{OrthoRemark6}) is zero for $kk'>0$, we obtain the orthogonal relation (\ref{OrthoScatt1}) of the scattering states.

\section{Green function $G^{(0,0)}(x,x';t)$}
\label{GreenFunctDeriv}

   In this appendix we show Eq. (\ref{GreenFunct2}). 

   By Eqs. (\ref{ScattState1}) and (\ref{ScattMatri3}) we note
%
\begin{widetext}
\begin{eqnarray}
   \Phi_{(j,k)}^{(0)}(x)\Phi_{(j,k)}^{(0)}(x')^{*}
   &=& \left|b_{(j,k)}^{(0)}\right|^{2} 
       \left[e^{ik(x-L)} + \frac{1+i\frac{\mu}{k}}{1-i\frac{\mu}{k}} e^{-ik(x-L)}\right]
        \left[e^{-ik(x'-L)} + \frac{1-i\frac{\mu}{k}}{1+i\frac{\mu}{k}} e^{ik(x'-L)} \right]
      \nonumber \\
   &=& 2\left|b_{(j,k)}^{(0)}\right|^{2} 
        \left\{\cos[k(x-x')] + \Real{\frac{1-i\frac{\mu}{k}}{1+i\frac{\mu}{k}} 
        e^{ik(x+x'-2L)}}\right\}
      \nonumber \\
   &=& \frac{4}{\pi}\frac{\cos[k(x-x')]
      +\frac{\left(1-\frac{\mu^{2}}{k^{2}}\right)
      \cos[k(x+x'-2L)]+2\frac{\mu}{k}\sin[k(x+x'-2L)]}{1+\frac{\mu^{2}}{k^{2}}}}
      {\left|N+1+i\frac{\lambda}{k}+\left(N-1+i\frac{\lambda}{k}\right)
      \frac{1+i\frac{\mu}{k}}{1-i\frac{\mu}{k}}e^{2ikL}\right|^{2}} 
\label{GreenFunct4}
\end{eqnarray}
\end{widetext}
where we used Eqs. (\ref{ScattCoeff1}), (\ref{ScattCoeff3}) and (\ref{ConstA1}). 
Here, $\Real{X}$ is the real part of $X$ for any complex number $X$.  
   It is important to note that the quantity (\ref{GreenFunct4}) is an even function of $k$, namely 
\begin{eqnarray}
   \Phi_{(j,-k)}^{(0)}(x)\Phi_{(j,-k)}^{(0)}(x')^{*} 
   = \Phi_{(j,k)}^{(0)}(x)\Phi_{(j,k)}^{(0)}(x')^{*} .
\label{EvenPhiPhiK1}
\end{eqnarray}
By inserting Eq. (\ref{GreenFunct4}) into Eq. (\ref{GreenFunct1}) for $n=n'=0$ and by noting Eq. (\ref{EvenPhiPhiK1}) we obtain Eq. (\ref{GreenFunct2}) with Eqs. (\ref{FunctTau1}) and (\ref{FunctF1}).

\section{Escape behaviors in the open boundary case} 
\label{OpenBoundaryCase}

   In this appendix we discuss asymptotic escape behavior of a particle in the open boundary case $\mu = 0$. 
   In this case, the function (\ref{FunctF1}) is represented as 
\begin{eqnarray}
   F(x,x';k) = \frac{4N}{\pi}\frac{\cos[k(x-L)]\cos[k(x'-L)]}{C(k)}
\label{FunctF4}
\end{eqnarray}
where $C(k)$ is given by 
\begin{eqnarray}
   C(k) &=& \left[N+1+(N-1)\cos(2kL)-\lambda\frac{\sin(2kL)}{k}\right]^{2} 
      \nonumber \\
   && \spaEq + \left[(N-1)\sin(2kL)+\lambda\frac{1+\cos(2kL)}{k}\right]^{2} .
   \nonumber \\ &&
\label{FunctC2}
\end{eqnarray}
In this case, the expansion of the function (\ref{FunctF4}) with respect to $k^{2}$ as in Eq. (\ref{GreenFunct3}) is different between the cases of $\lambda = 0$ and $\lambda \neq 0$, so we discuss these two cases separately below.

\subsection{Open boundary case with $\lambda \neq 0$} 

In the open boundary case $\mu = 0$ with nonzero potential barrier with $\lambda \neq 0$ at the junction $O$, the expansion coefficients $B_{\nu}(x,x')$, $\nu=0,1,\cdots$ of the function (\ref{FunctF4}) with respect to $k^{2}$, like in Eq. (\ref{FunctF2}), are concretely represented as 
%
\begin{widetext}
\begin{eqnarray}
   B_{0}(x,x') &=& 0,
      \\
   B_{1}(x,x') &=& \frac{N}{\pi\lambda^{2}} ,
      \\
   B_{2}(x,x') &=& -\frac{N}{2\pi \lambda^{2}}\left[(x-L)^{2}+(x'-L)^{2}
      +2\frac{N^{2}-2\lambda L-(\lambda L)^{2}}{\lambda^{2}}\right] ,
\end{eqnarray}
$\cdots$, concretely. By the Green function (\ref{GreenFunct3}) with these coefficients $B_{\nu}(x,x')$, $\nu = 0,1,\cdots$, the wave function (\ref{WaveFunct2}) for $t>0$ is expressed as  
\begin{eqnarray}
   \Psi^{(0)}(x,t) &=& 
       -\frac{N(1+i)\Theta_{0}}{2\sqrt{2\pi} \lambda^{2}} \frac{1}{\tau_{t}^{3/2}}
       +\frac{3N(1-i)}{8\sqrt{2\pi}\lambda^{2}}\left[ (x-L)^{2}\Theta_{0}+ \Theta_{2}
       +2\frac{N^{2}-2\lambda L-(\lambda L)^{2}}{\lambda^{2}}\Theta_{0}
       \right] \frac{1}{\tau_{t}^{5/2}}
       +\cdots
\label{WaveFunct4}
\end{eqnarray}
\end{widetext}
with $\Theta_{\nu}$ is defined by Eq. (\ref{FunctTheta1}).  

   By Eq. (\ref{WaveFunct4}) the survival probability $P(t)$ is represented asymptotically in time as 
\begin{eqnarray}
   P(t) \;\overset{t\rightarrow+\infty}{\sim}\;  \frac{2N^{2}\Xi_{0}L}{\pi\lambda^{4}}\left(\frac{m}{\hbar t}\right)^{3}  
\label{SurviAsym4}
\end{eqnarray}
with the constant $\Xi_{0}\equiv \Theta_{0}^{2}/\int_{0}^{L} dx \; \rho^{(0)}(x,0)$ 
determined by the initial wave function $\Psi^{(0)}(x,0)$ only. 
   Eq. (\ref{SurviAsym4}) leads to the relations
\begin{eqnarray}
   \lim_{t\rightarrow+\infty} \frac{P(t)}{P(t)|_{N=1}} &=& N^{2} ,
      \label{SurviPrope4} \\
   \lim_{t\rightarrow+\infty} \frac{P(t)}{P(t)|_{\lambda=1}} &=& \frac{1}{\lambda^{4}} 
\end{eqnarray}
for the $N$- and $\lambda$- independent initial wave function $\Psi^{(0)}(x,0)$. 
   Therefore, the $N$-dependence (\ref{SurviPrope4}) of the asymptotic survival probability for the open boundary case with $\lambda \neq 0$ is the same as that shown in Eq. (\ref{SurviPrope2}) for $\lambda L\neq -1$ in the fixed boundary case. 
   In contrast, the asymptotic survival probability is inversely proportional to $\lambda^{4}$ in Eq. (\ref{SurviAsym4}), while in the fixed boundary case with $\lambda L\neq -1$ it is inversely proportional to $(1+\lambda L)^{4}$ as shown in Eq. (\ref{SurviAsym1}).

   From Eqs. (\ref{LocalVeloc1}) and (\ref{WaveFunct4})  we also derive an asymptotic expression of the local velocity $v^{(0)}(x,t)$ in the finite region $\caL^{(0)}$ as   
\begin{eqnarray}
   v^{(0)}(x,t) \;\overset{t\rightarrow+\infty}{\sim}\; -\frac{3(L-x)}{t} .
\label{LocalVeloc4}
\end{eqnarray}
Therefore, the escape velocity (\ref{LocalVeloc2}) is given by 
\begin{eqnarray}
   V(t) \;\overset{t\rightarrow+\infty}{\sim}\; \frac{3L}{t} 
\label{AsympEscapVeloc2}
\end{eqnarray}
asymptotically in time, meaning that the escape velocity decays in power $\sim t^{-1}$ as in the fixed boundary case, but its value is three times as high as in Eq. (\ref{AsympEscapVeloc1}).

\subsection{Open boundary case with $\lambda =0$}

In the case of zero potential barrier with $\lambda =0$, from Eqs. (\ref{FunctF2}) and (\ref{FunctF4}) we derive the quantities $B_{\nu}(x,x')$, $\nu=0,1,\cdots$ as
%
\begin{widetext}
\begin{eqnarray}
   B_{0}(x,x') &=& \frac{1}{\pi N} ,
      \\
   B_{1}(x,x') &=& -\frac{1}{2\pi N}\left[(x-L)^{2}+(x'-L)^{2}-2L^{2}\frac{N^{2}-1}{N^{2}}\right] ,
\end{eqnarray}
$\cdots$, concretely, in the open boundary case $\mu = 0$. 
   By using these coefficients  $B_{\nu}(x,x')$, $\nu = 0,1,\cdots$ and Eqs. (\ref{WaveFunct2}) and (\ref{GreenFunct3}), the wave function $\Psi^{(0)}(x,t)$ for $t>0$ in $\caL^{(0)}$ is represented as 
\begin{eqnarray}
   \Psi^{(0)}(x,t) = 
       \frac{ (1-i)\Theta_{0}}{\sqrt{2\pi} N}\frac{1}{\tau_{t}^{1/2}}
       +\frac{1+i}{4\sqrt{2\pi} N}\left[ (x-L)^{2} \Theta_{0} 
       + \Theta_{2}-2L^{2}\frac{N^{2}-1}{N^{2}} \Theta_{0}\right] 
       \frac{1}{\tau_{t}^{3/2}}
       +\cdots
\label{WaveFunct5b}
\end{eqnarray}
\end{widetext}
with $\Theta_{\nu}$ defined by Eq. (\ref{FunctTheta1}).  

   By Eq. (\ref{WaveFunct5b}) the survival probability $P(t)$ is represented asymptotically in time as 
\begin{eqnarray}
   P(t) \;\overset{t\rightarrow+\infty}{\sim}\; \frac{2\Xi_{0}mL}{\pi N^{2}\hbar t} .
\label{SurviAsym2}
\end{eqnarray}
   Eq. (\ref{SurviAsym2}) shows that the survival probability $P(t)$ in the case of $\mu=0$ and $\lambda=0$ decays asymptotically in power $\sim t^{-1}$, as in Eq. (\ref{SurviAsym3}) for the fixed boundary case with $\lambda L =-1$. 
   Moreover, from Eq. (\ref{SurviAsym2}) we derive the relation with the same as Eq. (\ref{SurviPrope3}) for the $\lambda$ and $N$-independent initial wave function  $\Psi^{(0)}(x,0)$. 

   From Eqs. (\ref{LocalVeloc1}), (\ref{FunctTau1}) and (\ref{WaveFunct5b}) we derive the local velocity $v^{(0)}(x,t)$ in the finite region $\caL^{(0)}$, which is asymptotically the same as Eq. (\ref{LocalVeloc3}) in time. 
   Therefore, we obtain the same asymptotic escape probability $V(t)$ as Eq. (\ref{AsympEscapVeloc1}) for the fixed boundary case with $\lambda L\neq -1$.




\begin{thebibliography}{99}
\bibitem{Memo3} The local velocity (\ref{LocalVeloc1}) is defined only at the point where the particle  exists, i.e. at the point $x$ satisfying the condition $\Psi^{(n)}(x,t)\neq 0$.
\bibitem{Memo5} At the time $t$ satisfying the condition $\lim_{x\rightarrow +0}\Psi^{(n)}(x,t)= 0$ or $\Psi^{(0)}(L,t)= 0$, the boundary condition (\ref{BoundMomen1}) or (\ref{BoundMomen2a}) is not well defined, because here the local velocity $\lim_{x\rightarrow +0} v^{(n)}(x,t)$ or $v^{(0)}(L,t)$ cannot be introduced \cite{Memo3}. 
   For such a case, the conditions (\ref{BoundMomen1}) and (\ref{BoundMomen2a}) have to be replaced by the conditions $\lim_{x\rightarrow +0} \sum_{n=0}^{N} \caJ^{(n)}(x,t) = 0$ and $\caJ^{(0)}(L,t) = 0$ with $\caJ^{(n)}(x,t) \equiv (\hbar/m) \Imaginary{\Psi^{(n)}(x,t)^{*}\partial \Psi^{(n)}(x,t)/\partial x} = \rho^{(n)}(x,t) v^{(n)}(x,t)$, respectively.
\bibitem{Memo7}  
   In principle, the conditions (\ref{BoundMomen1}) and (\ref{BoundMomen2a}) allow the quantities $\mu$ and $\lambda$ in Eqs.  (\ref{BoundVeloc1}) and (\ref{BoundMomen2}) to depend on time $t$, respectively, but in this paper we consider only the cases in which both of $\mu$ and $\lambda$ are time-independent constants.
\bibitem{Memo1} 
   In general, open quantum network systems may have bound states with discretized energy spectrum, which are orthogonal to scattering states \cite{AS91}. 
   Effects of such bound states in escape phenomena are discussed in Ref. \cite{DN02}.
\bibitem{Memo2} 
   In this paper we do not prove the completeness of the scattering states. 
   Therefore, even if we introduce the initial states as the normalized one $\InPro{\Psi(0)}{\Psi(0)} =1$, the right-hand side of the initial state (\ref{InitiState1}) might not be normalized. 
   We should also note that we need a different normalization factor for the systems shown schematically in Figs. \ref{Fig1SystemA} and \ref{Fig1SystemB}. 
   By these reasons we do not assume the normalization of the initial wave function in this paper, and  instead of it we put the factor $1/\int_{0}^{L} dx \; \rho^{(0)}(x,0)$ in the right-hand side of Eq. (\ref{SurviProba1}) so that we can still impose the condition $P(0)=1$. 
   We do not need this kind of factor for Eq. (\ref{LocalVeloc2}), because the local velocity (\ref{LocalVeloc1}) is given from the ratio $\partial \Psi^{(n)}(x,t)/\partial x/\Psi^{(n)}(x,t)$ so that the prefactor of the wave function $\Psi^{(n)}(x,t)$ is cancelled out in the local velocity $v^{(n)}(x,t)$.
\bibitem{Memo4} The probability for particles to remain in the initially confined region has been called by a couple of different names, such as the survival probability \cite{MHC01,FKC01,BD07,ZB03,TS11}, the non-escape probability \cite{DN02,GMV07,M03,C11} and the quantum staying probability \cite{BG01}, etc.
\bibitem{Memo6} The initial condition expressed by Eqs. (\ref{InitiState2}) and (\ref{InitiState3}) satisfies the boundary condition (\ref{BoundPosit1}). 
   Although for this initial condition we cannot apply the boundary conditions (\ref{BoundMomen1}) and (\ref{BoundMomen2a}) because of $\Psi^{(n)}(+0,0)= 0$ or $\Psi^{(0)}(L,0)= 0$ \cite{Memo3}, the modified boundary conditions $\lim_{x\rightarrow +0} \sum_{n=0}^{N} \caJ^{(n)}(x,0) = 0$ and $\caJ^{(0)}(L,0) = 0$ \cite{Memo5} for such a case are satisfied because of $\lim_{x\rightarrow +0} \caJ^{(n)}(x,0) = 0$ and $\caJ^{(0)}(L,0) = 0$.
\end{thebibliography}
\end{document}